\documentclass[a4paper,11pt]{article}
\usepackage{pos}

\title{Application of Machine Learning to Identify Radio Pulses of Air Showers at the South Pole}
\manuallySeparateAuthors
\author[a,b]{Frank G.~Schroeder}
\author[a]{ and Abdul Rehman}
\author[c]{ for the IceCube Collaboration}

\affiliation[a]{Bartol Research Institute, Department of Physics and Astronomy, University of Delaware,\\
  Sharp Lab, 104 The Green, Newark DE, 19716, United States of America}
\affiliation[b]{Institute for Astroparticle Physics, Karlsruhe Institute of Technology (KIT),\\
Postfach 3640, 76021 Karlsruhe, Germany}
\affiliation[c]{Full author list at \href{https://icecube.wisc.edu/collaboration/authors/\#collab=IceCube&date=2024-08-15&formatting=web}{https://icecube.wisc.edu/collaboration/authors/}}

\emailAdd{fgs@udel.edu}
\emailAdd{arehman@udel.edu}

\abstract{Machine learning is a useful tool for identifying radio pulses from cosmic-ray air showers and for cleaning such pulses from radio background. 
This can lower the detection threshold and increase the accuracy for the pulse time and amplitude. We have trained Convolutional Neural Networks (CNNs) using CoREAS simulations and background recorded by a prototype station at the IceTop surface array at the South Pole and have applied them to air-shower measurements by this station. 
The station consists of 3 SKALA antennas and 8 scintillators, which are used to trigger the readout of the antennas upon a sixfold coincidence. Afterwards, the radio signal is filtered to the band of 70-350 MHz. 
By applying neural networks to search for radio signals in about four months of data, we find about five times more events than by a traditional method based on a signal-to-noise ratio cut after filtering for radio frequency interferences. 
Despite the lower threshold, the purity of the selected events seems to improve, and the angular resolution of the radio measurements does not deteriorate, which we have confirmed by a comparison of the reconstructed shower direction with IceTop. 
This analysis thus provides experimental confirmation that neural networks can indeed be used to clean air-shower radio signals from background and to lower the radio detection threshold of hybrid arrays combining particle and radio detectors.}

\FullConference{10th International Workshop on Acoustic and Radio EeV Neutrino Detection Activities (ARENA2024)\\
11-14 June 2024\\
The Kavli Institute for Cosmological Physics, Chicago, IL, USA\\}

\begin{document}
\maketitle

\section{Introduction}
The threshold of radio detection of cosmic-ray air showers is limited by external radio background, with Galactic noise and anthropogenic radio-frequency interferences (RFI) being the major backgrounds for most air-shower radio arrays \cite{Huege:2016veh,Schroder:2016hrv}.
Techniques to reduce the impact of background include interferometry \cite{ANITA:2010ect,LOPES:2021ipp,Allison:2018ynt,PierreAuger:2023opi}, the choice of an optimum frequency band \cite{BalagopalV:2017aan}, and computational methods, such as artificial neural networks. 
Pioneering work using either simulated background \cite{Erdmann:2019nie} or measured background by Tunka-Rex \cite{Bezyazeekov:2015rpa,Bezyazeekov:2021sha} showed that radio pulses from air showers can be detected and denoised using a convolutional neural network (CNN) of the autoencoder type. 
Such CNNs can be used for identifying air-shower radio pulses (classifier) and for reconstructing the true radio pulse by removing the impact of background on the pulse shape, amplitude, and arrival time (denoiser).

Building on this prior work, we have developed and trained classifier and denoiser CNNs for the a prototype station of an enhancement of IceTop \cite{IceCube:2012nn}, the surface array for the IceCube Neutrino observatory at the South Pole, which features three SKALA antennas \cite{7297231} triggered by four pairs of scintillation panels \cite{Shefali:2022kvp}. 
We used CoREAS simulations and background measured by these antennas to train CNNs and applied them to air-shower measurements with these antennas. 
Using the simultaneous measurement of the same showers by IceTop as a reference, we performed an experimental test of the performance of the CNNs and compared it to a traditional method of applying a threshold cut on the signal-to-noise ratio (SNR) after RFI filtering \cite{IceCube:2021qnf}.
We also show based on the simulations and measured background that the CNNs will enhance the accuracy of the measurement of the pulse power and arrival time in the individual antennas.

\begin{figure}[b]
\begin{center}
    \includegraphics[width=0.53\linewidth]{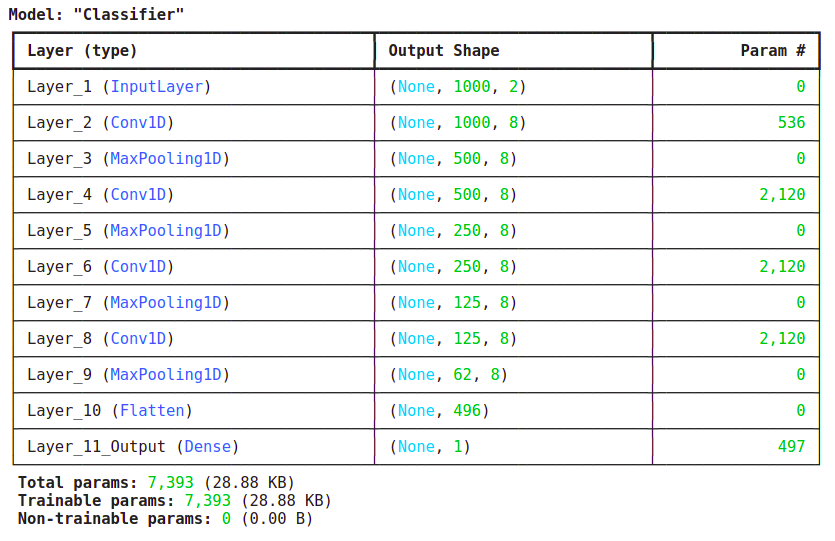}
    \hfill
    \includegraphics[width=0.46\linewidth]{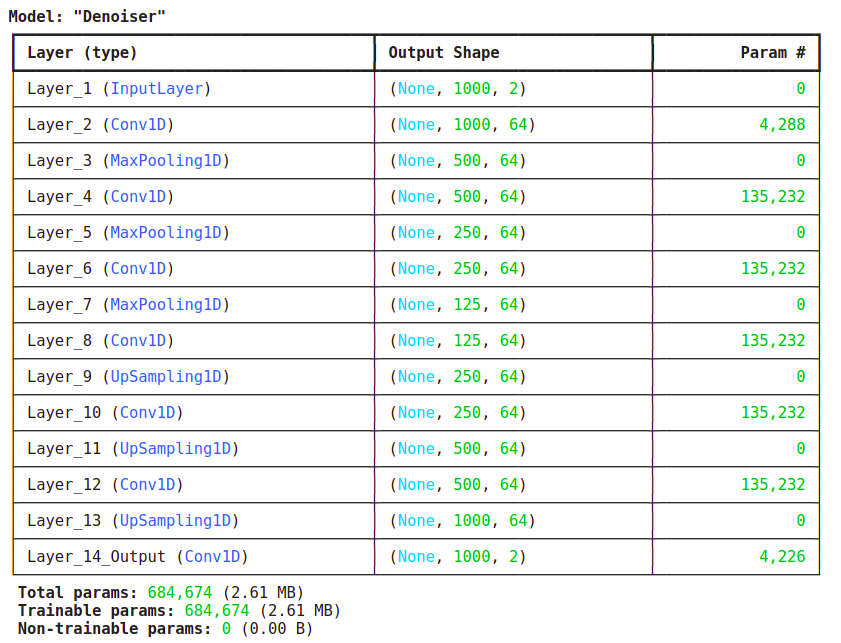}
    \caption{Parameter summary of the Classifier (left) and Denoiser (right) CNNs. The “None” in the Output Shape column indicates that the number is not pre-defined and represents the value of the input sample. This flexibility enables as many samples as desired for network training. The second and third numbers in the input layer represent the trace length and number of channels, respectively.}
    \label{fig:NetworkArch}
\end{center}
\end{figure}

\section{Data Set}
We have trained classifier and denoiser CNNs (see figure \ref{fig:NetworkArch}) using CoREAS \cite{Huege:2013vt} pulses and background measured by the same radio setup at the South Pole, to which we later apply the networks for the detection and reconstruction of air-shower radio pulses.

\begin{figure}[t]
\begin{center}
    \includegraphics[width=0.41\linewidth]{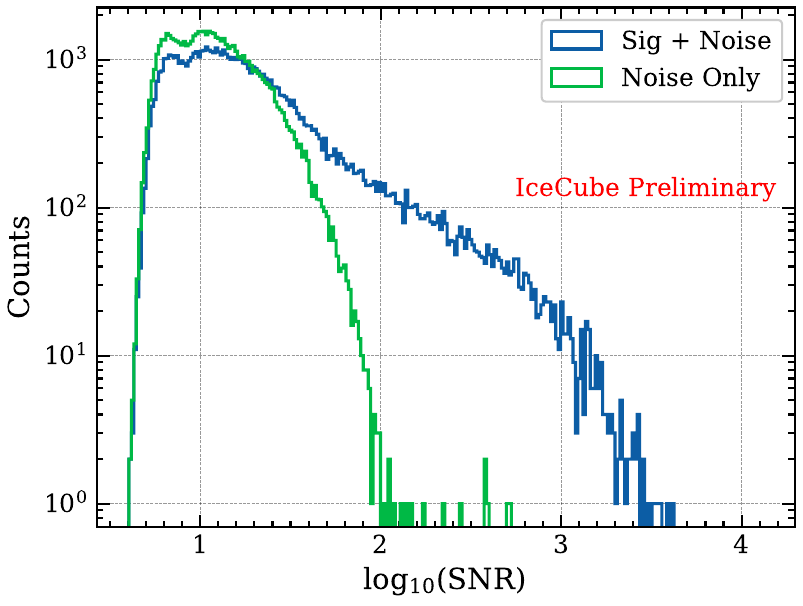}
    \hfill
    \includegraphics[width=0.48\linewidth]{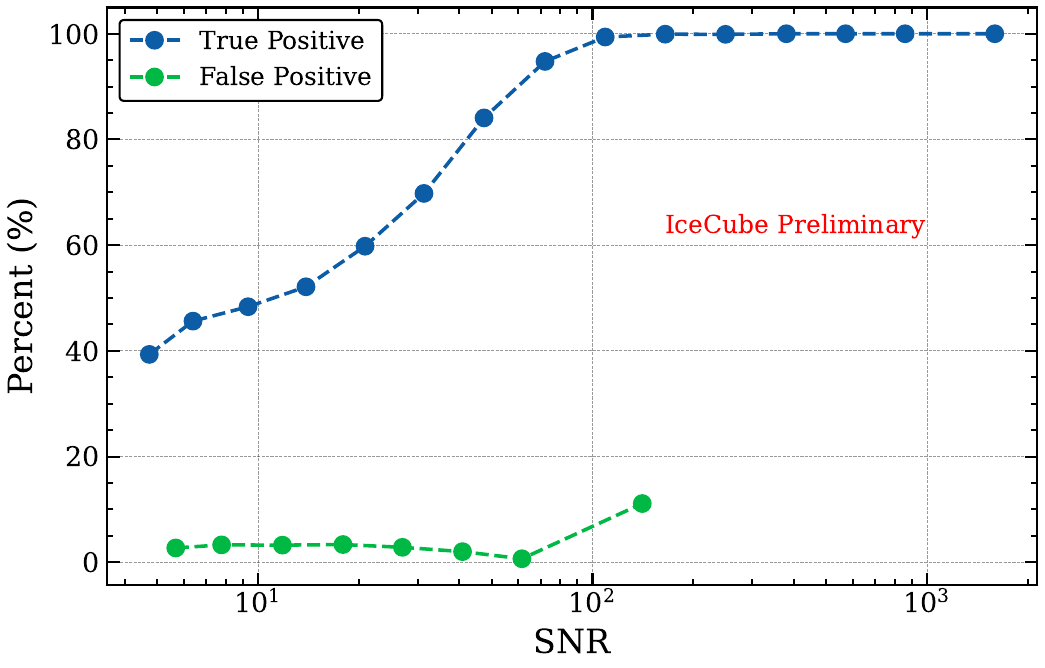}
    \caption{Left: Distribution of the signal-to-noise ratio (SNR) for measured background (Noise Only) and CoREAS pulses with background added (Sig + Noise) used for the training and testing of the networks. Right: Fraction of true positive and false positive events over SNR when using a threshold of $0.5$ on the output of the classifier network.}
    \label{fig:SnrAndTPFP}
\end{center}
\end{figure}

For the CoREAS simulation data set, we produced approximately $40,000$ CoREAS simulations in the energy range of $10^{16}-10^{18}\,$eV with random azimuth angles and zenith angles varying from $0^\circ$ to $71.6^\circ$. 
The simulations are filtered to $70-350\,$MHz and resampled to traces of $1000$ samples at $1\,$GSp/s, i.e., a length of $1\,$\textmu s.
The radio instrumentation response is applied to mimic measured signals \cite{IceCube:2022dcd}.
Measured background traces are chopped to the same trace lengths and added to the simulations, creating one data set of more than $160,000$ events for each of the three antennas, with each event consisting of a pair of traces corresponding to the two polarization channels.
Thereby, we train separate networks for each antenna using the background measured with the respective antenna. 
Figure~\ref{fig:SnrAndTPFP} (left) shows the distribution of the signal-to-noise ratio (SNR; see \cite{Schroder:2023sam}) of the data set for one antenna for illustration.
Using the Keras and TensorFlow software packages for the neural networks, the data set is split into training and test subsets. 
This proceeding presents the results obtained with the test data set, which meanwhile have been confirmed with an independent validation data set.

For the air-shower data set, we use $227,613$ scintillator-triggered events recorded during different periods from January to July 2022. 
The large majority of these events are at low energies, without any detectable radio signal expected. 
Using a traditional method of RFI filtering and a subsequent SNR threshold, we found $111$ events with signal in all three antennas and a reconstructed arrival direction consistent within $7^\circ$ with the independent IceTop reconstruction of the same events.
This traditional analysis described in more detail in reference \cite{MeghaARENA2024} serves as a baseline to evaluate the performance of the CNNs.

\begin{figure}[t]
\begin{center}
    \includegraphics[width=0.99\linewidth]{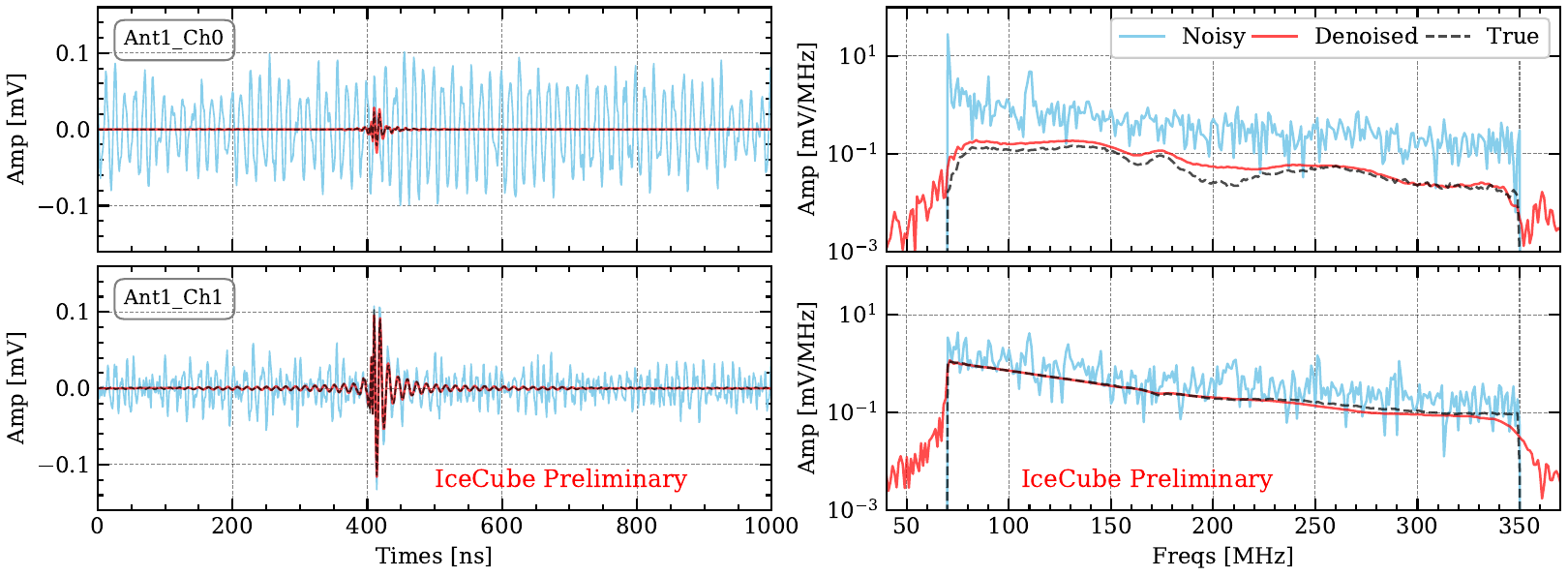}
    \caption{Example event illustrating the performance of the denoiser network on both polarization channels. In one channel, the radio pulse has an amplitude below the background level, but the network is able to recover the time of the pulse and the shape of the frequency spectrum even though the amplitude is too low. In the other channel, the pulse amplitude is higher than the noise level and the true pulse is fully recovered.}
    \label{fig:DenoExamples}
\end{center}
\end{figure}

\begin{figure}[t]
\begin{center}
    \includegraphics[width=0.49\linewidth]{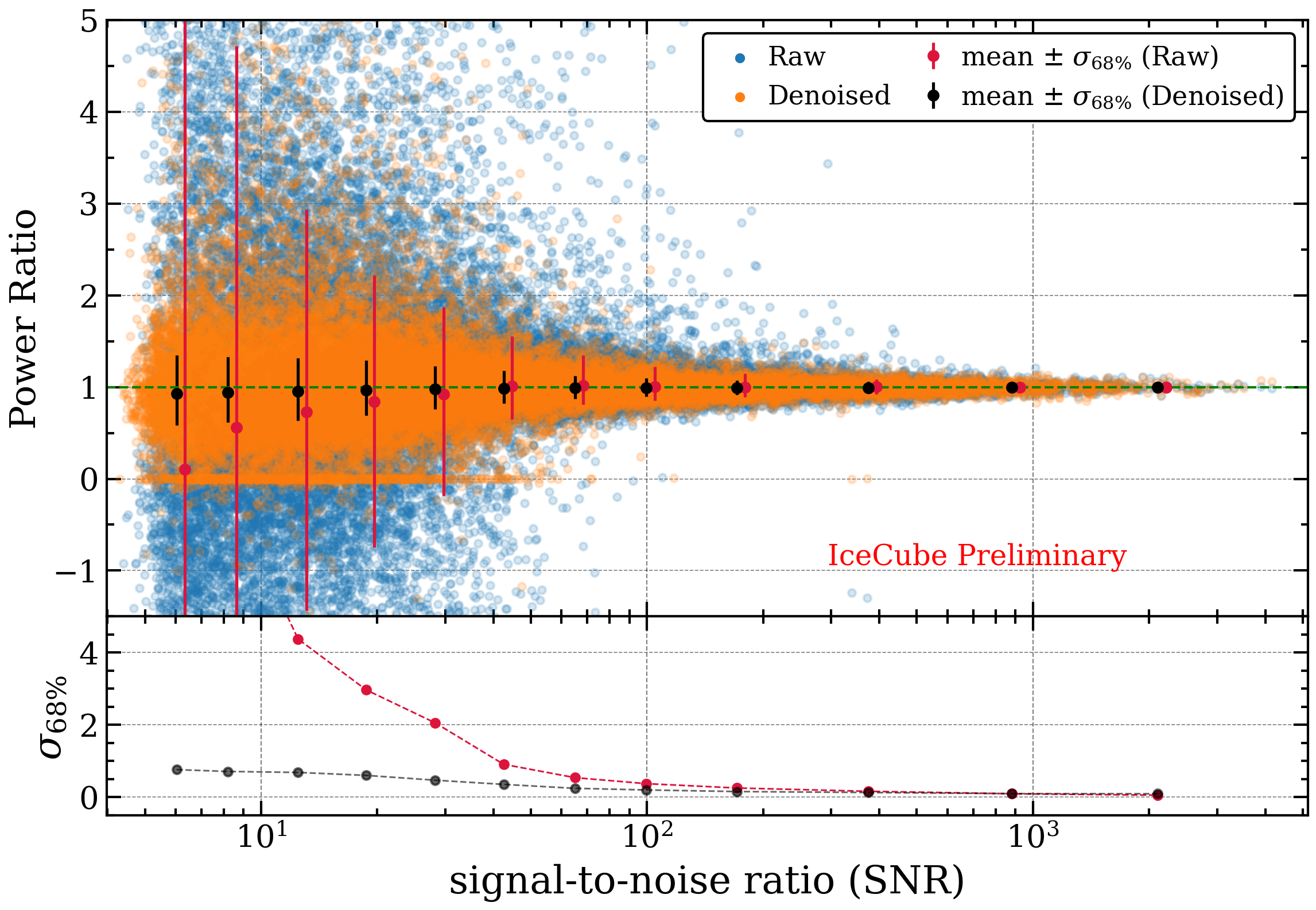}
    \hfill
    \includegraphics[width=0.49\linewidth]{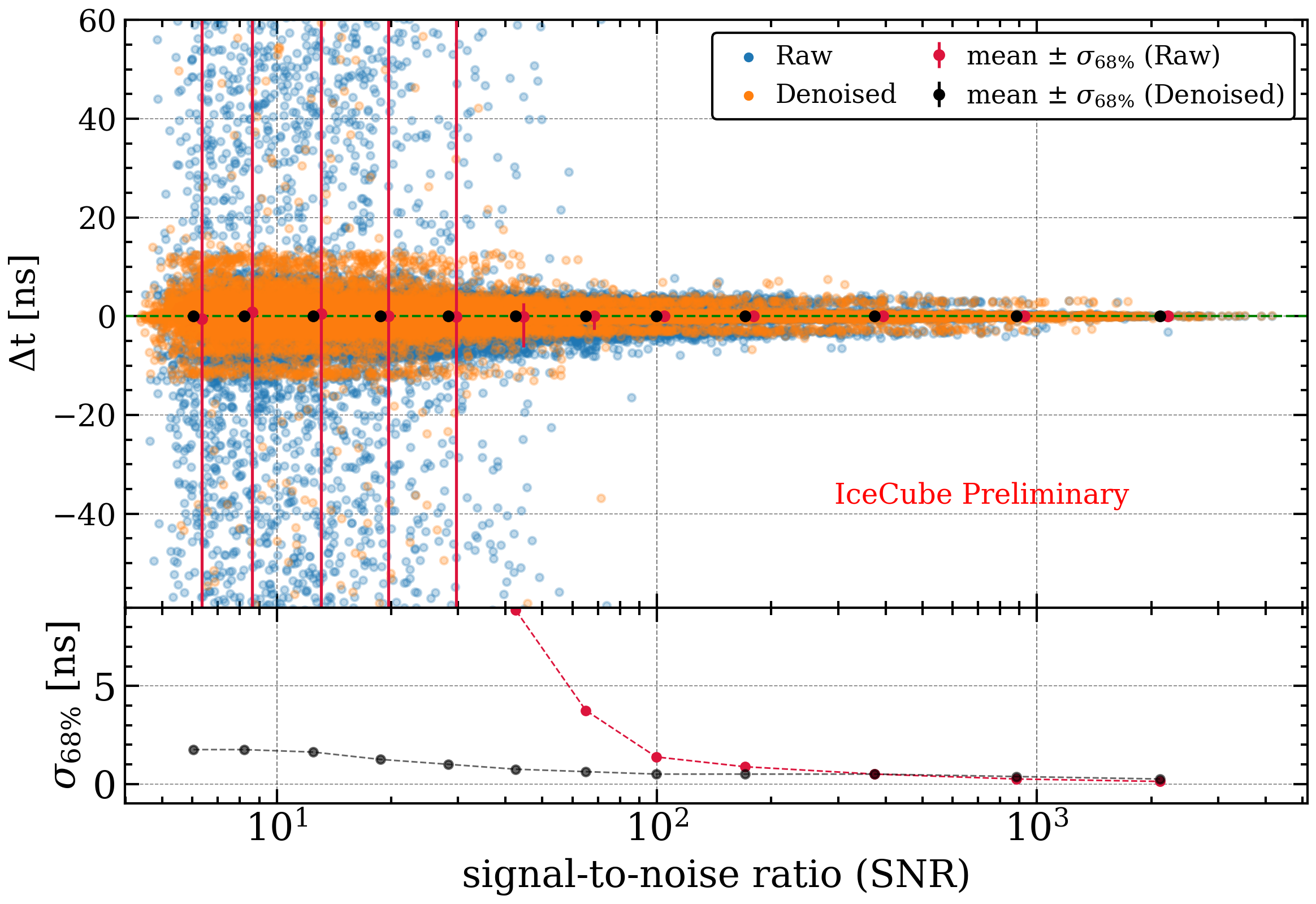}
    \caption{Ratio of the true and reconstructed power of the radio pulse (left) and difference of the peak pulse time $\Delta t$ over signal-to-noise ratio (SNR). The orange and blue markers in both plots represent the computed values for the raw (preprocessed but not denoised) and denoised waveforms, respectively. The red and black markers denote the mean values, with the bars indicating the $68\,\%$ containment (see \cite{rehman2021classification} for detailed description of both quantities). Negative power ratios before denoising are due to subtracting the background power from the pulse power, and 0 values of the denoiser are due to pulses not recognized. The large scatter of the values before denoising is due to background fluctuations mistaken for the air-shower pulse because we look for the highest peak in a pre-defined search window.}
    \label{fig:AccDenoiser}
\end{center}
\end{figure}

\section{Network Performance}
The performance of the classifier network has first been tested on the simulation data set with background. 
Figure~\ref{fig:SnrAndTPFP} (right) illustrates that at high SNR values almost all radio pulses are identified by the network, so there is no disadvantage to a traditional SNR cuts there. 
At low SNR values (including values below the mean SNR of background of $14.2$), a substantial fraction of the CoREAS radio pulses are correctly recognized, which would be rejected when applying a strict threshold in SNR, instead. 

The performance of the denoiser network is illustrated for one example event in Figure~\ref{fig:DenoExamples}.
Figure~\ref{fig:AccDenoiser} shows the performance for determining the pulse power, calculated as average power in a $100\,$ns interval around the pulse peak, and for the pulse peak time. 
Again, at high SNR values, there is no major difference to the classical approaches of determining the power of the radio pulse from the noisy trace (after subtracting the average power of the noise) and of using the peak time of the noise pulse. 
However, at intermediate and low SNR values, the denoiser network clearly improves the accuracy of both, the pulse power and peak time.

%\begin{figure}[h!]
%\begin{center}
%    \includegraphics[width=0.98\linewidth]{figs/Example_AirShowerEvent.png}
%    \caption{An example air shower event recorded by the three antennas shown in light blue. The denoised waveforms are shown in red.}
%    \label{fig:exampleEvent}
%\end{center}
%\end{figure}

\begin{figure}[t]
\begin{center}
    \includegraphics[width=0.48\linewidth]{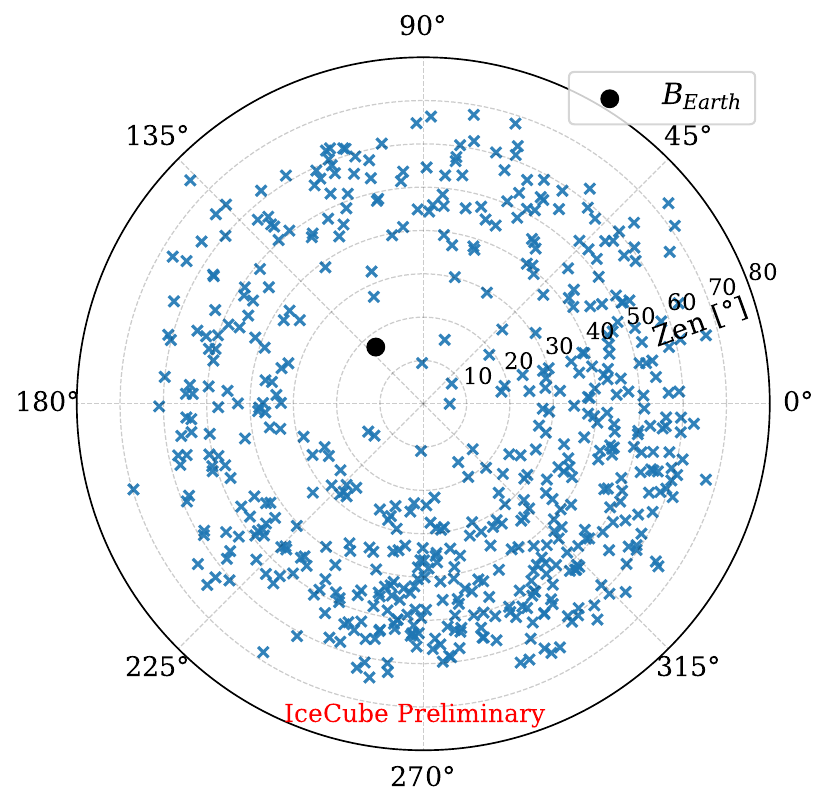}
    \hfill
    \includegraphics[width=0.51\linewidth]{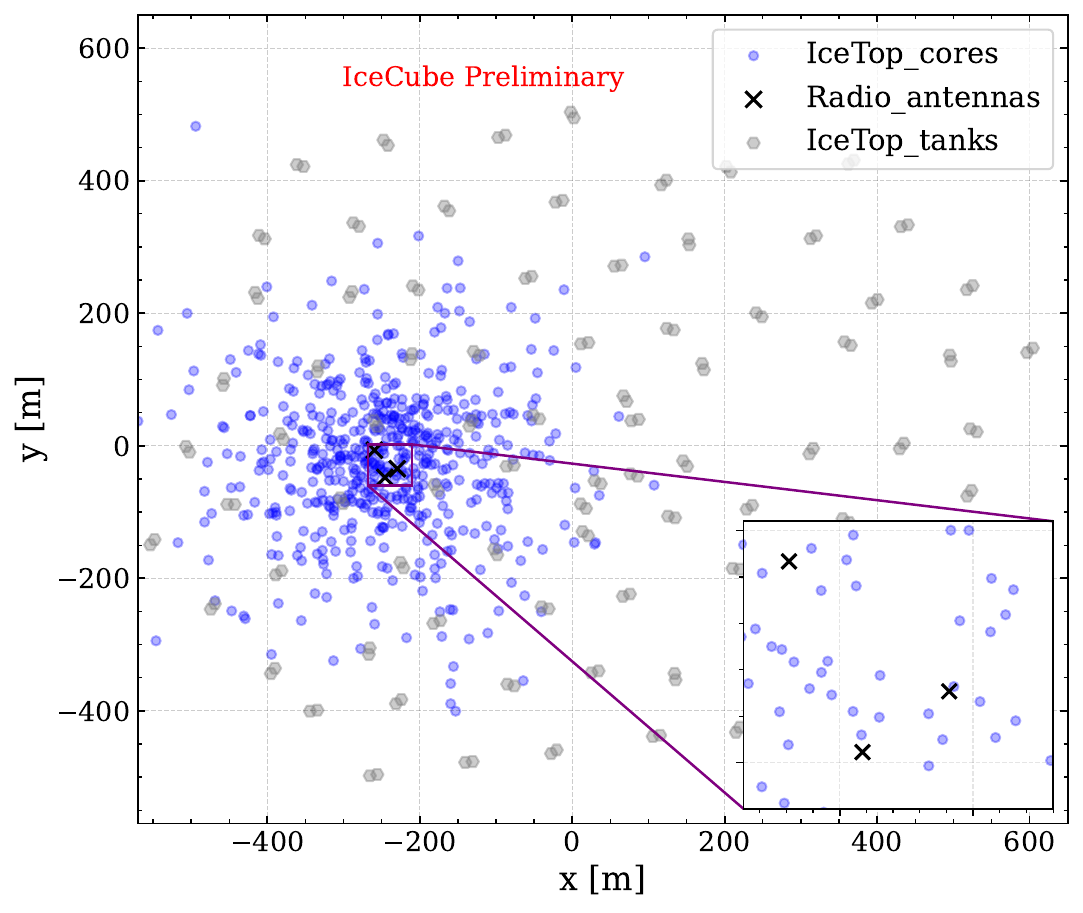}
    \caption{Arrival directions (left) and shower cores (right) of the air-shower events measured by the prototype station at the South Pole and identified and reconstructed using the CNNs.}
    \label{fig:candidateDist}
\end{center}
\end{figure}

\section{Performance on Air-Shower Measurements}
The classifier and denoiser networks have been applied to the scintillator-triggered radio events of the prototype station at the South Pole and compared to coincident IceTop measurements of the same air showers.
We have applied a combined cut on the radio pulse amplitude and output score of the classifier to search for air-shower signals, where the threshold for the classifier score starts at $0.6$ at low pulse amplitudes and ramps down to $0$ for very large pulses clearly above the average background level. 
These values have been chosen such that $95\,\%$ of background traces from the fixed-rate trigger are rejected, so the performance can be compared to the traditional method of an SNR threshold, which is also set to reject $95\,\%$ of background traces in individual antenna channels.

With the CNNs, $608$ events pass the cut in all three antennas compared to $341$ with the SNR method.
However, these are contaminated by false-positive detections, as we only require a $95\,\%$ rejection in the individual antennas.
To reject most of these false positives, we compare a plane-wave reconstruction of the arrival direction to the independent direction reconstruction of IceTop. 
Then, a data-driven cut of a maximum opening angle of $7^\circ$ is applied as an additional selection criterion\footnote{For the SNR-threshold analysis in reference \cite{MeghaARENA2024}, we choose $5^\circ$, which reduced the expected contamination from $4$ to $2$ false-positive events ($<1$ event for the CNN method in both cases); all conclusions hold for either choice.}, which leaves $554$ events for the CNN method and $111$ events for the SNR method (of which $103$ events were found with both methods).

\begin{figure}[t]
\begin{center}
    \includegraphics[width=0.63\linewidth]{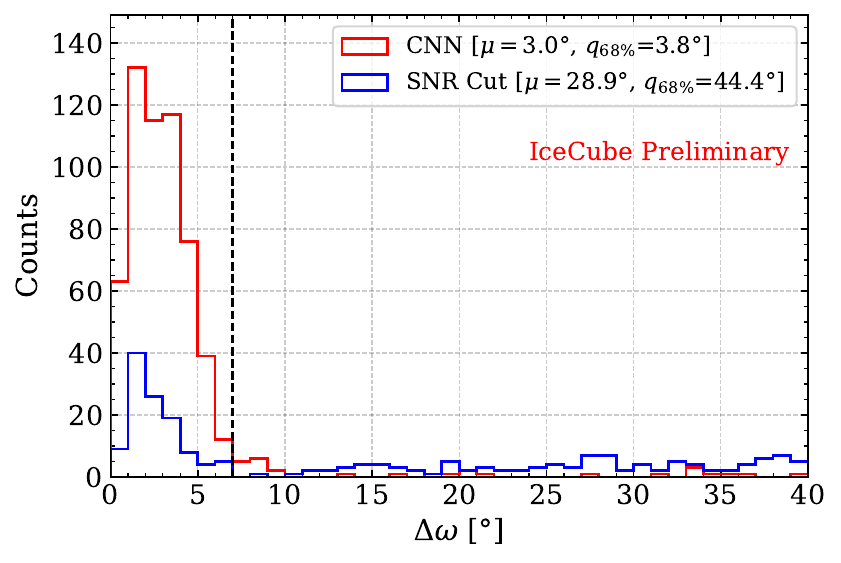}
    \caption{Angular difference $\Delta\omega$ between the IceTop and radio direction reconstructions for the CNN and SNR events. The dashed line indicates the $7^\circ$ cut motivated by the distribution of the CNN events.}
    \label{fig:DOmegaComp}
\end{center}
\end{figure}

Figure~\ref{fig:candidateDist} shows the distribution of the arrival directions and core distributions of the events identified with the CNN method. 
As expected due to the size of the radio footprint and the dominant geomagnetic emission mechanism, most events are within $200\,$m distance from the antennas and at large geomagnetic angles (only few events are at zenith angles $>60^\circ$ because of the limited efficiency of the scintillators used for triggering). 
Figure~\ref{fig:DOmegaComp} compares the reconstructed radio and IceTop arrival directions of the events for both methods. 
The peak at low angular differences provides additional experimental confirmation that we measured real air-shower events.
The angular distribution also indicates that the CNN method provides a larger and purer sample than the SNR method: there are almost five times more events below $7^\circ$, i.e., more air showers detected in radio, and four times less events at larger opening angles, which indicates less false positive detections contaminating the data sample. 
A comparison to IceTop's energy estimator confirm that many of the additional events of the CNN sample are at lower energy than events in the SNR sample, which provides further confirmation that the CNNs lower the detection threshold for air-shower radio pulses.

\section{Conclusion}
We have presented CNNs for the identification and denoising of air-shower radio pulses and applied these CNNs to measurements of a surface prototype station at IceCube featuring three elevated antennas. 
CoREAS simulations with measured background used for training and testing indicate that the CNNs increase the accuracy on the pulse power and arrival time, in particular, at low and intermediate signal-to-noise ratios.
Consequently, the CNNs lower the detection threshold which leads to the identification and successful direction reconstruction of about five times more air-shower events with the CNNs compared to a traditional method using an SNR threshold. 
As an additional benefit, the CNNs also reduce the number of false-positive detections of background fluctuations, thus, not only increasing the statistics, but also the purity of radio measurements of air showers.

\bibliographystyle{ICRC}
\bibliography{bibliography}

\section*{Acknowledgement}

{
The authors gratefully acknowledge the support from several agencies and institutions. In addition to those acknowledged with the \href{https://icecube.wisc.edu/collaboration/authors/#collab=IceCube&date=2024-08-15&formatting=web}{full authorlist}, we specially acknowledge:\\
This research was supported in part through the use of Information Technologies (IT) resources at the University of Delaware, specifically the high-performance computing resources. This project has received funding from the European Research Council (ERC) under the European Union’s Horizon 2020 research and innovation programme (grant agreement No 802729). The work has also been supported by NASA EPSCoR. 
}

\end{document}